\documentstyle[aps,floats,prb,epsfig]{revtex}
\begin{document}


\newlength{\plstm}
\settowidth{\plstm}{$+$}
\addtolength{\plstm}{-.07\plstm}

\newcommand{\plustimes}{$+\hspace{-\plstm}\times$}



\twocolumn[\hsize\textwidth\columnwidth\hsize\csname @twocolumnfalse\endcsname

\title{\bf Dynamical properties of Au from tight-binding 
molecular-dynamics simulations}

\author{F. Kirchhoff,$^1$ M. J. Mehl,$^2$ N. I. Papanicolaou,$^3$
D. A. Papaconstantopoulos,$^2$ and F. S. Khan$^1$}

\address{$^1$ Department of Electrical Engineering, Ohio State
University, Columbus, OH 43210}
\address{$^2$ Center for Computational Materials Science, Naval
Research Laboratory, Washington DC 20375-5000}
\address{$^3$ Department of Physics, Solid State Division,
University of Ioannina, P.O.~Box 1186, GR-45110 Ioannina, Greece}

\date{\today}

\maketitle

\begin{abstract}
We studied the dynamical properties of Au using our previously
developed tight-binding method.  Phonon-dispersion and
density-of-states curves at T=0~K were determined by computing the
dynamical-matrix using a supercell approach.  In addition, we
performed molecular-dynamics simulations at various temperatures to
obtain the temperature dependence of the lattice constant and of the
atomic mean-square-displacement, as well as the phonon
density-of-states and phonon-dispersion curves at finite
temperature.  We further tested the transferability of the model to
different atomic environments by simulating liquid gold.  Whenever
possible we compared these results to experimental values.
\end{abstract}
]
%

\section{Introduction}
\label{sec:intro}

Over the last two decades atomistic simulations have become an
increasingly important tool for modeling in many areas of
condensed-matter physics and material science. The most challenging
problem in computer-based nano-scale simulations of real materials
is to find an accurate and transferable model for the atomic
interactions that reproduces the energetic and electronic properties
of the material.  A whole hierarchy of models for atomic
interactions have been developed, ranging from simple empirical
potentials to sophisticated first-principles calculations based on
density-functional-theory (DFT).  Although DFT methods are very
accurate and have been successfully applied to the study of a broad
range of materials and systems, they are computationally very
demanding. Even with today's state-of-the-art computers, DFT
simulations with more than 100 atoms are challenging.  Empirical
potentials, on the other hand, are less demanding and have been used
to simulate systems with millions of atoms. This advantage is
however to be weighed against a loss in accuracy and
transferability.

Several empirical potential methods have been used in the past to
simulate metallic systems: the embedded-atom method, the
effective-medium theory, Finnis-Sinclair potentials and the
second-moment approximation to the tight-binding model.\cite{sma}
The decade has seen the emergence of a method that lies between
first-principles and empirical potentials: the so-called
tight-binding (TB) molecular-dynamics method. It is more accurate
than the empirical potential methods because it explicitly describes
the electronic-structure of the system. TB is roughly three orders
of magnitude faster than DFT based methods due to the much smaller
size of the secular equation, which makes the N$^3$ issue more
tolerable.  The TB method has been used to study a broad range of
materials.\cite{tbreview}

Recently the NRL group proposed an alternative formulation of the TB
method, which was shown to work well for transition
metals,\cite{papamehl} simple metals,\cite{simplemetals} and
semi-conductors.\cite{sc} This approach has been successful in
determining static properties such as structural energy differences,
elastic constants, vacancy formation energies and surface energies.

Although static calculations are very useful for determining many
fundamental properties of materials, such calculations are limited
to properties at T=0~K.  Most problems in real materials involve
processes that occur at finite temperature.  The purpose of the
present work is to demonstrate that our TB model can successfully be
applied to the study of the dynamical and finite temperature
properties of a representative material, gold.  Our previous TB
parmetrization of gold\cite{papamehl} was highly successful in
predicting structural properties.  We have improved upon this
parametrization in this paper.  This provides us with an ideal test
case for demonstrating the power of the method.

We tested our TB parameters by calculating the elastic properties
and comparing to first-principles calculations and experiment.  We
also found the the phonon-dispersion curves and density-of-states
(DOS) at T=0~K by calculating the dynamical-matrix using a supercell
method.\cite{dynmat} In addition, we performed molecular-dynamics
(MD) simulations at various temperatures to obtain the temperature
dependence of the lattice constant and of the atomic
mean-square-displacement, as well as the electronic and phonon DOS
and the phonon-dispersion curves at finite temperatures and a
simulation of the liquid phase.  Whenever possible we compare these
results to experimental data.

\section{Technical details}
\label{sec:method}

\noindent{\em Fitting procedure for Au}
\vspace{3mm}

Details about our TB model can be found in Ref.~\cite{papamehl}.  In
this paper we used a new TB parametrization for Au,\cite{goldpar}
which works well even at very small interatomic distances.  The
parameters of the model are fitted to reproduce data from DFT
calculations: band structures and total energy as a function of
volume for {\em fcc}, {\em bcc} and simple cubic ({\em sc})
structures.  In the present case the database included ten (10) {\em
fcc} structures, six (6) {\em bcc} structures, and five (5) {\em sc}
structures.  The calculations were performed using the general
potential Linearized Augmented Plane Wave (LAPW)
method,\cite{andersen75,singh86} using the
Perdew-Wang\cite{perdew92} parametrization of the Local Density
Approximation.\cite{kohn65} In addition care was taken to include
energies at very small volumes (down to 60\% of the equilibrium
volume) in the fitting database. This turned out to be very
important in order to have parameters that could be used in the wide
range of interatomic distances that occur during MD simulations.
Finally it should be stressed that no experimental data is used to
determine the parameters of the model.

\vspace{3mm}
\noindent{\em DoD-TBMD code}
\vspace{3mm}

Except as noted, the results presented in this paper were obtained
using the DoD-Parallel Tight-Binding Molecular-Dynamics (TBMD) code
developed as part of the Computational Chemistry and Materials
Science (CCM) contribution to the Common HPC Software Support
Initiative (CHSSI).  This program was written with the goal of
performing molecular-dynamics simulations of metallic systems.
Although initially written to run with our TB
Hamiltonian,\cite{papamehl} this code is in fact model
independent.\cite{othermodels} The electronic structure is
calculated using either a O(N$^3$) method such as diagonalization,
or by using an O(N$^2$) method called the Kernel Polynomial Method
(KPM).\cite{kpm,cub} The code has been written for both scalar and
parallel computers.  The parallel parts of the code have been written
using a message-passing programming model relying on the MPI library
to deal with communications.\cite{moreoncode}

\vspace{3mm}
\noindent{\em Simulation details}
\vspace{3mm}

To compute the dynamical-matrix we used an {\em fcc} supercell of
1331 atoms, obtained by replicating a primitive {\em fcc} cell 11
times along the three primitive lattice vectors.  Periodic boundary
conditions are applied throughout this work.  In the MD simulations
the system consists of an {\em fcc} supercell of 343 atoms, obtained
by replicating a primitive {\em fcc} cell 7 times along the three
primitive lattice vectors.  The Brillouin-zone (BZ) is sampled using
the $\Gamma$-point.  We checked that this is a reasonable
approximation even for a metal: the lattice constant, bulk modulus
and elastic constants obtained from the 343 atom supercell and
$\Gamma$-point sampling are within 10\% of the values obtained using
a primitive cell and a well converged k-point set.  The MD
simulations were started with atoms arranged on an {\em fcc} lattice
and random velocities drawn from a Boltzmann distribution for a
temperature $2T$.  The MD simulation was performed in the
micro-canonical ensemble, so at equilibrium the temperature of the
lattice averaged to $T$, and the ``potential energy'' of the system
was raised by an amount $3/2~Nk_BT$, where $N$ is the number of
atoms and $k_B$ is Boltzmann's constant.  The equations of motion
were integrated using the Verlet algorithm and a time step of
$\Delta{t} = 2$~fs, giving a total-energy conservation within
$\Delta{E}/E = 10^{-5}$.  The system was equilibrated at the desired
temperature for 1500 time steps (3~ps).  Typically another 1500
additional steps were performed to calculate time averages.  The
finite size of the simulation cell and the finite number of time
steps in the simulation implies that the instantaneous temperature
of the cell, computed from the kinetic energy, fluctuated around
some average temperature, which was not necessarily the target
temperature.  For the simulations conducted at a target temperature
of 300K we found that the average temperature after the system
reached equilibrium was 301K, with a standard deviation of 8K.  At
600K, we found the average temperature to be 604K with standard
deviation 20K, and at 1200K we found 1212K and 113K, respectively.

For the computation of the temperature dependence of the lattice
constant we used a 64 atom {\em fcc} supercell and sampled the BZ
with four $k$-points.  For each volume and temperature we ran a
Langevin dynamics simulation\cite{langevin} for 2.5~ps, using a time
step of 5~fs and a friction parameter $\gamma$=0.05~fs$^{-1}$.


\section{Results and discussion}
\label{sec:results}
\noindent{\em Equation of State}\vspace{3mm}

\begin{figure}
\epsfig{file=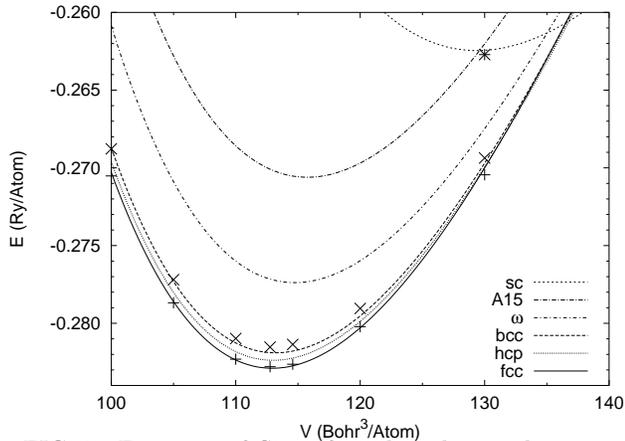,width=3.35in}
\caption{Equation of State for selected crystal structures of gold
using the tight-binding parameters discussed in the text.  All
coordinates are relaxed at each volume.  The points are the
first-principles LAPW energies used in the fit.  From bottom to top,
the ordering of the structures is {\em fcc} (LAPW symbol $+$), {\em
hcp}, {\em bcc} (LAPW symbol $\times$), hexagonal $\omega$, A15, and
simple cubic ({\em sc}, LAPW symbol \plustimes).}
\label{fig:eos}
\end{figure}

In Fig~\ref{fig:eos} we present the energy versus volume curves for
a selection of crystal structures, calculated using the {\tt static}
TB code.\cite{static} The LAPW total energies for the {\em fcc},
{\em bcc}, and {\em sc} structures, which were used in the fit, are
also plotted on the graphs.  We find that the TB Hamiltonian
predicts that the equilibrium {\em fcc} structure has a lower energy
than any other structure yet tested, consistent with experiment and
first-principles calculations, and confirming the robustness of the
Hamiltonian.

\vspace{3mm}\noindent{\em Elastic properties}\vspace{3mm}

The elastic properties of bulk {\em fcc} Au were calculated using
our TB parameters by means of the standard finite strain
method\cite{mehl90,mehl95} and the {\tt static} code.  The results
are shown in Table I, along with comparisons to experimental
data\cite{simmons} and the results of first-principles LAPW
calculations.  The latter calculations were also performed using the
Perdew-Wang LDA parametrization.\cite{perdew92} The TB calculations
reproduce the LAPW results very well and are in good agreement with
the experimental data.

\vspace{3mm}\noindent{\em Phonons at T=0~K}\vspace{3mm}

We determined the phonon dispersion curves and density-of-states
(DOS) of {\em fcc} Au by computing the dynamical matrix. This was
achieved by using a large supercell, in our case containing 1331
atoms, and calculating the forces on all atoms in response to the
displacement of the atom at the origin. Provided this displacement
is small enough it is possible to construct the real-space
dynamical-matrix using finite differences and compute the
dynamical-matrix by a Fourier series.\cite{dynmat}

\begin{figure}
\epsfig{file=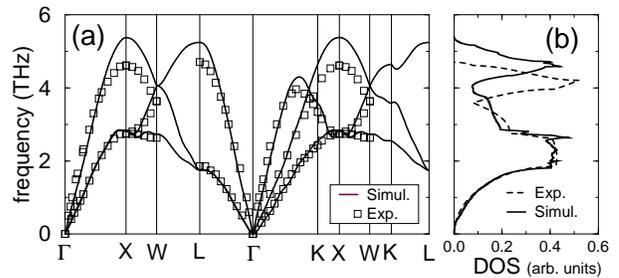,width=3.35in}
\caption{(a) Phonon-dispersion curves for Au at T=0~K, plotted along
high-symmetry directions in the BZ.  Lines are spline fits to the
theoretical TB data, open squares are experimental data
points.\protect\cite{lynn73} (b) Phonon density-of-states (DOS) for
Au at T=0~K.  The full line is the theory result using our TB model,
while the dotted line is a fit to the experimental
values.\protect\cite{lynn73} The dispersion curves and DOS were
calculated from the dynamical-matrix computed using an {\em fcc}
supercell containing 1331 atoms.}
\label{fig:phonons0K}
\end{figure}

The high-symmetry direction phonon dispersion curves for Au at T=0~K
are shown in Fig.~\ref{fig:phonons0K}(a) together with experimental
data.\cite{lynn73} The overall structure of the dispersion curves is
well reproduced.  The low frequency transverse modes are in
excellent agreement with experiment.  The longitudinal
higher-frequency modes, however, are systematically too high close
to the BZ edge.

The phonon DOS is presented in Fig.~\ref{fig:phonons0K}(b) together
with experimental results. The DOS has two main peaks. From 0 to
3.5~THz the theoretical DOS reproduces the experimental data very
well. In contrast in the region 4-5~THz the position of the high
frequency peak in the theoretical curve is overestimated by about
0.5~THz compared to experiment, consistent with the discrepancy in
the frequency of the high-frequency longitudinal modes noted above.

\begin{table}
\caption{\label{tab:elast}Bulk modulus and elastic constants (in
GPa) for Au computed using our TB model compared to the results
of LAPW calculations and experimental data.  All calculations are
performed at the experimental room-temperature volume, the measured
elastic constants are taken from the compilation of Simmons and
Wang.\protect\cite{simmons}}
\begin{tabular}{lccccc} 
     & B   & C$_{11}$-C$_{12}$ & C$_{11}$ & C$_{12}$ & C$_{44}$ \\
\hline
TB   & 181 & 21 & 195 & 174 & 40 \\
LAPW & 182 & 27 & 200 & 173 & 33 \\
Exp. & 169 & 30 & 189 & 159 & 42 \\
\end{tabular}
\end{table}
We checked that the observed discrepancies are not an artifact of
the supercell method used to calculate the dynamical-matrix. In
particular we made sure that an 11$\times$11$\times$11 supercell is
large enough to converge the dynamical-matrix.  A first check is
given by the slopes of the dispersion curves as $k$$\rightarrow$0,
which are related to the elastic constants.  We found that the
slopes were consistent with our computed elastic constants (see
Table \ref{tab:elast}).  Another check is to compute the frequency
of BZ-edge phonons using the frozen-phonon
method\cite{frozenphonons} for a 2 or 4 atom unit cell, using the
{\tt static} code to calculate the total energies.  The computed
BZ-edge phonon frequencies are shown in Table
\ref{tab:frozen_phonons}.  They are in perfect agreement with the
frequencies derived from the dispersion curves obtained using the
dynamical-matrix method (see Fig.~\ref{fig:phonons0K}), hence
confirming the accuracy of the latter approach.  Finally using the
first-principles LAPW method we calculated the phonon frequencies at
$X$ and $L$ and found very good agreement with experiment as shown
in Table~\ref{tab:frozen_phonons}.  We therefore conclude that the
overestimate of our calculated longitudinal phonon frequencies near
the BZ-edge is a shortcoming of our TB parameters.  An obvious
approach to overcome this problem and improve the agreement with
experiment would be to include the LAPW frequencies at $X$ and $L$
in our fitting database.  Nevertheless we note that our TB results
for the dispersion curves present a substantial improvement over
results obtained using the second-moment approximation to
TB.\cite{sma}
\begin{table}
\caption{\label{tab:frozen_phonons}Selected phonon frequencies (in
THz) at high symmetry points in the Brillouin-zone.  The calculated
TB and LAPW frequencies were obtained with the frozen phonon method
in cells with 2 or 4 atoms.}
\begin{tabular}{lcccccc} 
&
X(L) & X(T) & L(L) & L(T) & W(L) & W(T) \\ 
\hline
TB   & 5.29 & 2.87 & 5.35 & 1.91 & 2.66 & 4.01 \\
LAPW & 4.43 &      & 4.53 &      &      &      \\ 
Exp.\protect\cite{lynn73} & 4.60 & 2.72 & 4.69 & 1.85 & 2.63 & 3.62 \\
\end{tabular}
\end{table}

\vspace{3mm}\noindent{\em Electronic Density of States at finite
temperature}\vspace{3mm}

\begin{figure}
\epsfig{file=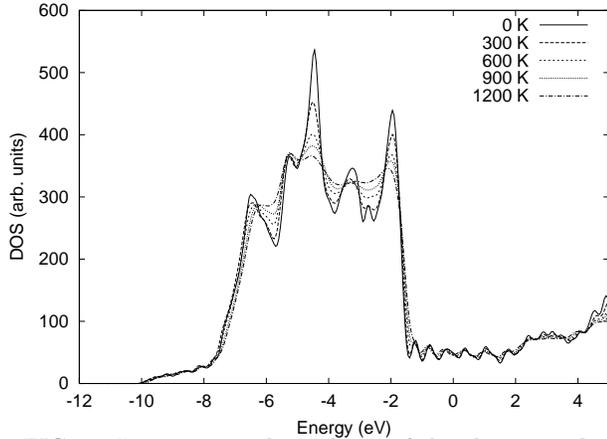,width=3.35in}
\caption{Temperature dependence of the electronic density of states
of gold, calculated using the eigenvalues generated by the TBMD
code, and averaged over ten time steps, as outlined in the text.}
\label{fig:tedos}
\end{figure}

The TBMD code computes all of the eigenvalues of the system at each
time step, making it simple to determine electronic properties such
as the density of states as a function of temperature.  In
Fig.~\ref{fig:tedos} we show the electronic DOS at several
temperatures.  For each temperature, we saved the eigenvalues for
ten different time steps.  The DOS was then calculated assuming a
Fermi distribution, and the resulting DOS were averaged.  We see
that the dominant effect of increasing temperature is to reduce the
peaks in the electronic DOS spectrum.

\vspace{3mm}\noindent{\em Phonons at finite temperature}\vspace{3mm}

\begin{figure}
\epsfig{file=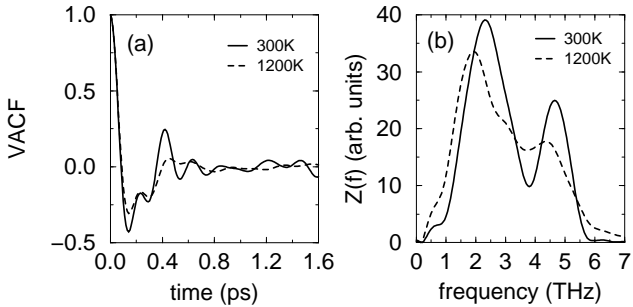,width=3.35in}
\caption{(a) Velocity-velocity auto-correlation functions (VACF) of
Au calculated from molecular-dynamics simulations at 300~K and
1200~K using our TB model. (b) Finite-temperature phonon
spectral-density (PSD) $Z(f)$ for Au at 300~K and 1200~K.  The PSD
was computed by Fourier-transform of the VACF shown in (a).}
\label{fig:vac}
\end{figure}

We determined the phonon dispersion curves and spectral-density of
{\em fcc} Au at finite temperature by performing MD simulations.  In
Fig.~\ref{fig:vac}(a) we show the velocity-velocity auto-correlation
functions (VACF) obtained from the MD simulation at 300~K and
1200~K.  We see that increasing temperature damps out the
oscillations in the VACF.

The finite-temperature phonon spectral-density (PSD) can be obtained
from the Fourier-transform of the VACF,\cite{vac} as shown in
Fig.~\ref{fig:vac}(b).  The PSD has the two well defined peaks,
consistent with the DOS at T=0~K.  The position of the peaks is also
in agreement with the theoretical data at T=0~K.  The limited
resolution in the finite-temperature PSD, due to the short length of
the MD simulation, makes a detailed comparison with experiment
difficult.  It should also be pointed out that the PSD is
proportional to the phonon DOS only in an harmonic solid; one should
therefore be cautious when comparing the PSD to the phonon DOS, in
particular at high temperature.  We believe this may explain the
difference in height between the two peaks in the PSD, in contrast
to the T=0~K phonon DOS, where both peaks have about the same
height.  Comparison of the PSD at 300 and 1200~K clearly reveals the
effect of temperature: a clear shift of phonon frequencies to lower
values and a broadening of the peaks in the PSD.

\begin{figure}
\epsfig{file=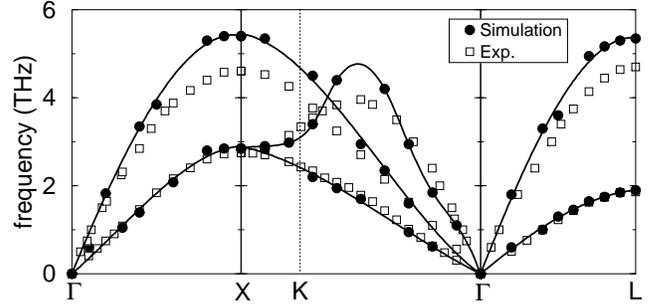,width=3.35in}
\caption{Finite-temperature phonon dispersion curves along the
high-symmetry directions in the BZ, for Au at 300~K calculated using
molecular-dynamics simulations based on our TB model.  Filled
circles are theoretical data, lines are polynomial fits to
theoretical data, open squares are experimental data
points.{\protect \cite{lynn73}} The dispersion curves were computed
by Fourier-transform of the time dependent wave-dependent VACF (see
text).}
\label{fig:disp300K}
\end{figure}

To compute the phonon dispersion curves we calculate the PSD from
the Fourier-transform of the velocity- and position-dependent
auto-correlation function.  Details of this computational procedure
can be found elsewhere.\cite{spec} The dispersion curves along
high-symmetry directions in the BZ at 300~K are reproduced in
Fig.~\ref{fig:disp300K}.  The comparison with experimental data
reveals the same discrepancies as in the T=0~K dispersion curves:
the high frequency longitudinal modes are overestimated close to the
BZ-edge.

We determined the temperature dependence of phonon frequencies by
performing MD simulations at 300, 600, 900 and 1200~K where for each
temperature we fix the volume at the experimental value.  In
Table~\ref{tab:phononsoft} we show the frequency of selected BZ-edge
phonons as a function of temperature.  These frequencies were
calculated as described above for the 300~K case.
\begin{table}
\caption{\label{tab:phononsoft}Selected phonon frequencies (in THz)
at high symmetry points in the Brillouin-zone, calculated as a
function of temperature (T in Kelvin) from MD simulations using our
TB model.}
\begin{tabular}{lcccc} 
    T (K) & X(T) & X(L) & L(T) & L(L) \\
\hline
  300 & 2.85 & 5.40 & 1.90 & 5.35 \\ 
  600 & 2.82 & 5.35 & 1.80 & 5.22 \\ 
  900 & 2.77 & 5.28 & 1.70 & 5.18 \\ 
 1200 & 2.65 & 5.20 & 1.60 & 5.15 
\end{tabular}
\end{table}

As expected the frequency of phonons decreases with temperature.
For the modes we computed the change was of the order of
0.2-0.3~THz, the transverse mode at L exhibiting the largest
decrease.  Part of this variation is probably due to increase in
volume as temperature increases.  It should be noted that using MD
to compute the phonon dispersion curves at finite temperature can be
particularly useful in systems where a particular crystal structure
is unstable at T=0~K ({\em bcc} Ti is one example) and where the
dynamical-matrix method will predict unstable phonon modes.  MD
simulation may also be useful to determine the vibrational
properties of systems for which an experimental study may be
difficult, e.g. clusters or nano-crystals.

\vspace{3mm}\noindent{\em Thermal expansion}\vspace{3mm}

To determine the theoretical thermal expansion coefficient $\alpha$ 
we use the following definition for $\alpha$:
\begin{equation}
\label{eq:alpha1}
\alpha = \frac{1}{3B}\left(\frac{\partial{P}}{\partial{T}}\right)_V.
\end{equation}
This definition requires the calculation of the pressure as a
function of temperature for a fixed volume.  We perform MD
simulations at 300, 600, 900 and 1200~K, keeping the volume fixed at
the experimental value at room temperature (lattice constant
$a~=~4.08$~\AA).  For each temperature we selected 10 independent
configurations from the trajectories generated by the MD and
computed the instantaneous pressure.  We found that 10
configurations per temperature were enough to get the average
pressure with an error margin of $\sim$5\%.  If, in
Eq.~\ref{eq:alpha1}, we assume the pressure varies linearly as a
function of temperature and if for $B$ we use the theoretical value
of the bulk modulus at T=0~K and at the experimental volume, we get
$\alpha=11\times 10^{-6}$~$\mbox{K}^{-1}$. This underestimates the
experimental value of 14$\times 10^{-6}$~$\mbox{K}^{-1}$ at
300~K.\cite{handbook}

An alternative definition of $\alpha$ is given by:
\begin{equation}
\label{eq:alpha2}
\alpha = \frac{1}{3V}\left(\frac{\partial{V}}{\partial{T}}\right)_P.
\end{equation}
To check the calculation of $\alpha$ based on Eq.~\ref{eq:alpha1} we
computed $\alpha$ using this latter definition. This requires MD
simulations for several volumes (typically 3 or 4) for a given
temperature.  For each volume $V$ we compute the average pressure
$P$.  The equilibrium volume at each temperature is found by
interpolating $P(V)$ to find the volume that gives zero pressure.
We performed this procedure at 300, 600, 900 and 1200~K to find
$V(T)$.

\begin{figure}
\epsfig{file=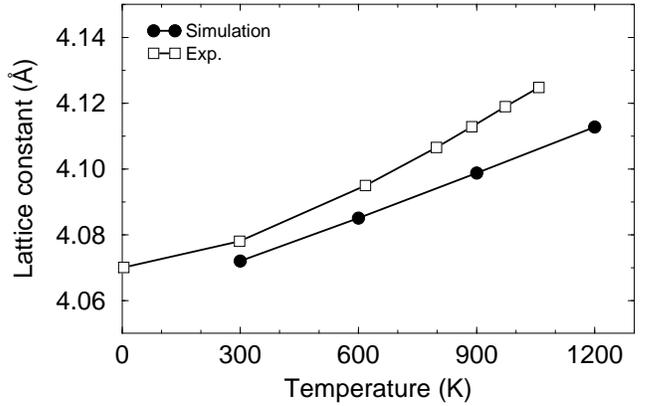,width=3.35in}
\caption{Lattice constant of Au as a function of temperature.  The
black circles are the results of the molecular-dynamics simulations
using our TB model, open squares are the experimental data.{\protect
\cite{pearson}}}
\label{fig:aot}
\end{figure}

In Fig.~\ref{fig:aot} we show the lattice constant as a function of
temperature as derived from the simulations, compared to
experimental results.\cite{pearson} The overall agreement with
experiment is good given that the theoretical data is well within
1\% of experiment in the temperature range we simulated.

From Fig.~\ref{fig:aot} we can see that it is reasonable to assume
that the volume varies linearly as a function of T. So by using
Eq.~\ref{eq:alpha2}, we get $\alpha = 11 \times
10^{-6}$~$\mbox{K}^{-1}$, in agreement with our previous estimate
based on Eq.~\ref{eq:alpha1}.

\begin{figure}
\epsfig{file=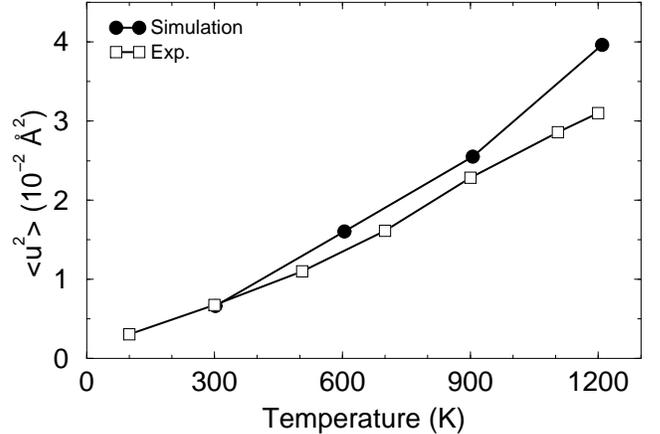,width=3.35in}
\caption{Mean-square displacement of Au as a function of
temperature.  The filled circles are the results of the
molecular-dynamics simulations using our TB model, empty squares are
the experimental points.{\protect\cite{owen}}}
\label{fig:msd}
\end{figure}

\vspace{3mm}\noindent{\em Mean-square displacement}\vspace{3mm}

We used the atomic positions generated by the MD simulations
performed for several temperatures at the corresponding experimental
lattice constants to compute the atomic mean-square displacement
(MSD).  In Fig.~\ref{fig:msd} we compare the temperature dependence
of our computed MSD with experimental data.\cite{owen} The agreement
with experiment is excellent up to 900K, at higher temperatures the
theoretical MSD gets larger than the experimental values.  Again it
should be noted that the results of our calculation are in much
better agreement with experiment compared to previous work using the
second-moment-approximation to TB.\cite{sma}

\vspace{3mm}\noindent{\em Liquid Au}\vspace{3mm}

\begin{figure}
\epsfig{file=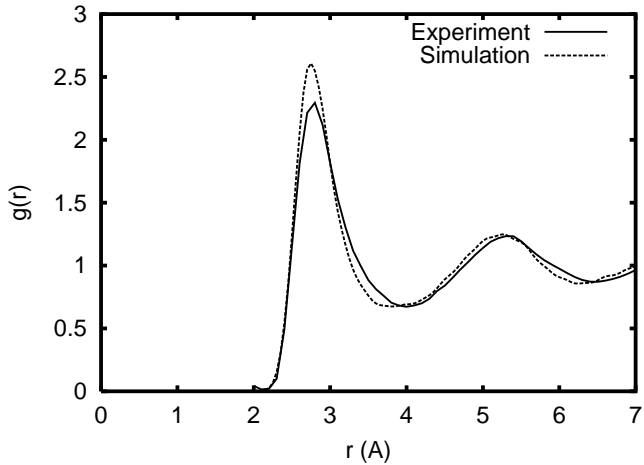,width=3.35in}
\caption{Pair correlation function $g(r)$ of liquid gold at 1773~K.
The dotted line is the result of molecular-dynamics simulations
using our TB model, the full line is the experimental
result.\protect\cite{waseda}}
\label{fig:gofr}
\end{figure}

To further test the transferability of our model to different atomic
environments we studied the liquid phase of Au.  The simulation was
performed with 200 atoms in a periodic {\em fcc} unit cell.  The
density of the sample was chosen to be equal to the experimental
value of 16.746 g cm$^{-3}$ at 1773~K.\cite{steinberg1974} One
thousand (1,000) MD steps were used to equilibrate the system.
Statistical averages of structural properties were computed from
data collected from the next two thousand (2,000) MD steps.  The
radial distribution function $g(r)$ obtained from our simulation
(shown in Fig.~\ref{fig:gofr}) is found to be in very good agreement
with experimental data.\cite{waseda}

\begin{figure}
\epsfig{file=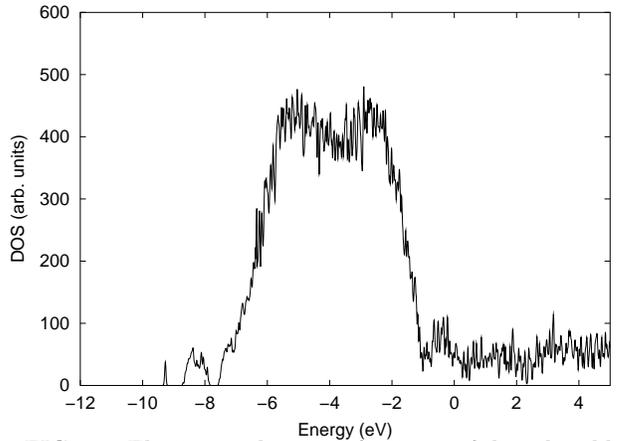,width=3.35in}
\caption{Electronic density of states of liquid gold at T~=~1773K,
using the same method as in Fig.~\protect\ref{fig:tedos}.}
\label{fig:t1773dos}
\end{figure}

We also calculated the electronic DOS of liquid gold at 1773~K,
using the same procedure as in the solid.  The result is shown in
Fig.~\ref{fig:t1773dos}.  While the overall shape of the DOS,
including the width, is similar to Fig.~\ref{fig:tedos}, the high
temperature and loss of symmetry has destroyed most of the peak
structure.  However, two new peaks appear at low energies, probably
due to the lack of periodicity in the liquid.

\section{Conclusions}
\label{sec:concl}

We presented results of simulations of bulk {\em fcc} Au using our
tight-binding model.  Our TB Hamiltonian was used to compute the
elastic constants of bulk Au, which were in very good agreement with
the results of LDA calculations and experimental data.  Using a
supercell method to compute the dynamical-matrix we determined the
phonon-dispersion curves and phonon density-of-states of Au at
T=0~K.  Our calculated dispersion curves are in good agreement with
experimental data, except for a tendency to overestimate the
frequency of longitudinal modes close to the Brillouin-Zone edge.
We checked that this discrepancy is not a consequence of the method
used to calculate the phonon frequencies.  In addition, we performed
molecular-dynamics simulations at various temperatures to compute
the phonon density-of-states and phonon-dispersion curves at finite
temperature.  The molecular-dynamics simulation were also used to
obtain the temperature dependence of the lattice constant and of the
atomic mean-square-displacement.  Both quantities were found to be
in good agreement with experimental data.  Finally, we performed an
MD simulation of the liquid phase of Au and obtained a radial
distribution function in very good agreement with experiment.  We
believe these results demonstrate that our TB model, using
parameters generated by the same procedure,\cite{tbpar} can
successfully be applied to the study of dynamical and finite
temperature properties of other metals.

\section*{Acknowledgments}
MJM and DAP are supported by the U. S. Office of Naval Research.
NIP acknowledges partial support from NATO grant CRG-940118.  The
development of the {\tt static} and {\tt DoD-TBMD} codes was
supported in part by the U. S. Department of Defense Common HPC
Software Support Initiative (CHSSI).  This work was supported in
part by a grant of HPC time from the DoD HPC Center, for
computations on the IBM SP2 and SGI Origin at the Aeronautical
Systems Center, Wright-Patterson Air Force Base, Dayton, OH.



\begin{thebibliography}{99}

\bibitem{sma} G. C. Kallinteris, N. I. Papanicolaou,
G. A. Evangelakis, and, D. A. Papaconstantopoulos, Phys. Rev. B {\bf
55}, 2150 (1997) and references therein.

\bibitem{tbreview}C. M. Goringe, D. R. Bowler, and E. Hernandez,
Rep. Prog. Phys. {\bf 60}, 1447 (1997); M. J. Mehl and
D. A. Papaconstantopoulos, {\em Computational Materials Science},
ed. C. Fong, World Scientific, Singapore, 1998.

\bibitem{papamehl}D. A. Papaconstantopoulos, M. J. Mehl,
Phys. Rev. B {\bf 54}, 4519 (1996).

\bibitem{simplemetals} S. H. Yang, , M. J. Mehl, and
D. A. Papaconstantopoulos, Phys. Rev. B {\bf 57}, R2013 (1998).

\bibitem{sc}D. A. Papaconstantopoulos, M. J. Mehl, S. C. Erwin and
M. R. Pederson, {\em Tight-binding approach to computational
material science}, edited by P. E. A. Turchi, A. Gonis and
L. Colombo, MRS symposia proceedings 491, Materials Research
Society, Pittsburgh, 1998, p. 221.

\bibitem{dynmat} See for example G. J. Ackland, M. C. Warren and,
S. J. Clark, J. Phys.: Condens. Matter {\bf 9}, 7861 (1997)
and references therein.

\bibitem{goldpar}The gold parameters used in this paper are
available from the authors, or at \\
{\tt http://cst-www.nrl.navy.mil/bind/au\_par\_99}.

\bibitem{andersen75}O. K. Andersen, {Phys. Rev. B} {\bf 12}, 3060
(1975).

\bibitem{singh86}D. Singh, H. Krakauer, and C. S. Wang, 
Phys. Rev. B {\bf 34}, 8391 (1986).

\bibitem{perdew92}J. P. Perdew and Y. Wang, Phys. Rev. B 
{\bf 45}, 13244 (1992).

\bibitem{kohn65}W. Kohn and L. J. Sham, Phys. Rev. {\bf 140},
A1133 (1965).               

\bibitem{othermodels} We have tested the code with a variety of TB
models found in the literature. Adding a new model requires only
minor changes to the code.

\bibitem{kpm} R. N. Silver, H. Roeder, A. F. Voter and J. D. Kress,
J. Comput. Phys. {\bf 124}, 115 (1996); A. F. Voter, J. D. Kress
and R. N. Silver, Phys. Rev. B {\bf 53}, 12733 (1996).

\bibitem{cub} In the present work the size of the simulated systems
is small enough that O(N$^3$) methods are faster than 
O(N$^2$) KPM methods.

\bibitem{moreoncode}More details about the DoD-Parallel
Tight-Binding Molecular-Dynamics code can be found at the following
URL: {\tt http://cst-www.nrl.navy.mil/bind/dodtb/}.

\bibitem{langevin} 
N. Binggeli and J. R. Chelikowsky, Phys. Rev. B {\bf 50}, 11764
(1994).

\bibitem{static}For more information about the {\tt static} TB code,
see\\
{\tt http://cst-www.nrl.navy.mil/bind/static/}.

\bibitem{mehl90}M. J. Mehl, J. E. Osburn, D. A. Papaconstantopoulos,
and B. M. Klein, Phys. Rev. B {\bf 41}, 10311 (1990); {\em
erratum}, Phys. Rev. B {\bf 42}, 5362 (1991).

\bibitem{mehl95}M. J. Mehl, B. M. Klein, and D. A.
Papaconstantopoulos, ``First-Principles Calculation of Elastic
Properties,'' in {\em Intermetallic Compounds: Principles and
Applications}, ed. J. H. Westbrook and R. L. Fleischer (London: John
Wiley \& Sons Ltd., 1995), Vol. 1, Ch. 9.

\bibitem{simmons}
G. Simmons and H. Wang, {\em Single Crystal Elastic Constants and
Calculated Aggregate Properties: A Handbook}, 
2$^{nd}$ Edition, (MIT Press, Cambridge, MA, 1971).

\bibitem{lynn73} J. W. Lynn, H. G. Smith and R. M. Nicklow,
Phys. Rev. B {\bf 8}, 3493 (1973).

\bibitem{frozenphonons} K.-M. Ho, C. L. Fu, B. N. Harmon, W. Weber and 
D. R. Hamann, Phys. Rev. Lett. {\bf 49}, 673 (1982).

\bibitem{vac} The Fourier transform of the VACF gives the spectral
density which is proportional to the phonon density-of-states in
an harmonic solid.

\bibitem{spec} N. I. Papanicolaou, I. E. Lagaris, and G. A.
Evangelakis, Surf. Sci. {\bf 337}, L819 (1995).

\bibitem{handbook} Y. S. Touloukian, {\em A physicist's desk
reference: The second edition of the physics vade mecum}, ed.\
H. L. Anderson, American Institute of Physics, New York, 1989,
p.~345.

\bibitem{pearson} W. B. Pearson, {\em A handbook of Lattice Spacings
and Structures of Metals and Alloys}, Pergamon, New York, 1958.

\bibitem{owen} E. A. Owen and R. W. Williams, Proc.~R.~Soc.~London,
Ser. A {\bf 188}, 509 (1947).

\bibitem{steinberg1974} D. J. Steinberg, Met. Trans. {\bf 5}, 1341
(1974).

\bibitem{waseda}Y. Waseda, {\em Structure of Non-Crystalline
Materials} (McGraw-Hill, New York, 1980).

\bibitem{tbpar}Parameters for other
elements\cite{papamehl,simplemetals,sc} are available at {\tt
http://cst-www.nrl.navy.mil/bind/}.

\end{thebibliography}
\end{document}